\documentclass[prd,aps,twocolumn,10pt,superscriptaddress]{revtex4-1}
\usepackage[utf8]{inputenc}
\usepackage{siunitx}
\usepackage{xcolor}
\usepackage{graphicx}
\graphicspath{{images/}}
\usepackage{amsmath}
\usepackage{hyperref}

\newcommand{\TEM}[2]{\ensuremath{{\rm HG}_{#1#2}}}
\newcommand{\figref}[1]{Fig.~\ref{fig:#1}}
\newcommand{\SQZ}[1]{\ifcase #1 \or SQZ1\or SQZ2\fi}
\newcommand{\Soo}{\Smn{00}}
\newcommand{\Loo}{\ensuremath{\text{LO}_{00}}}
\newcommand{\Shom}{\Smn{mn}}

\newcommand{\Smn}[1]{\ensuremath{\text{S}_{#1}}}
\newcommand{\unc}[1]{\sigma_{#1}}
\newcommand{\var}[1]{\unc{#1}^2}
\begin{document}

\title{Mitigating mode-matching loss in nonclassical laser interferometry}
\author{Sebastian Steinlechner}
\author{Niels-Ole Rohweder}
\author{Mikhail Korobko}
\affiliation{Institut f\"ur Laserphysik und Zentrum f\"ur Optische Quantentechnologien der Universit\"at Hamburg,\\%
Luruper Chaussee 149, 22761 Hamburg, Germany}
\author{Daniel T\"oyr\"a}
\author{Andreas Freise}
\affiliation{School of Physics and Astronomy and Institute of Gravitational Wave Astronomy, University of Birmingham, Birmingham B15 2TT, United Kingdom}
\author{Roman Schnabel}
\affiliation{Institut f\"ur Laserphysik und Zentrum f\"ur Optische Quantentechnologien der Universit\"at Hamburg,\\%
Luruper Chaussee 149, 22761 Hamburg, Germany}

\begin{abstract}
    Strongly squeezed states of light are a key technology in boosting the sensitivity of interferometric setups, such as in gravitational-wave detectors. However, the practical use of squeezed states is limited by  optical loss, which reduces the observable squeeze factor. Here we experimentally demonstrate that introducing squeezed states in additional, higher-order spatial modes can significantly improve the observed nonclassical sensitivity improvement when the loss is due to mode-matching deficiencies. Our results could be directly applied to gravitational-wave detectors, where this type of loss is a major contribution.
\end{abstract}
\maketitle

Nonclassical, squeezed states of light have progressed from being just peculiar quantum states to a quantum-mechanical technology that increases the sensitivity of interferometric measurements without the need for increasing laser powers. This is a remarkable feature with application in several areas \cite{schnabel_squeezed_2017}.

For example, gravitational-wave detectors already operate with extremely high laser powers that are limited both by engineering difficulties of the laser sources as well as the induced thermal load on the interferometer mirrors \cite{beausoleil_model_2003,ramette_analytical_2016}. While an increase in laser power by a factor of 10 (corresponding to a $10^{3/2} \approx 32$ increase in the observable universe at shot-noise limited detection frequencies) seems extremely challenging, the same improvement could be achieved with the injection of squeezed light with a noise-reduction factor of \SI{10}{dB}. Such strongly squeezed states are now routinely produced at near-infrared wavelengths \cite{vahlbruch_observation_2008,eberle_quantum_2010,stefszky_balanced_2012,schonbeck_13_2018,vahlbruch_detection_2016}. In 2010, GEO\,600 was the first gravitational-wave detector to employ squeezed light input \cite{the_ligo_scientific_collaboration_gravitational_2011,grote_first_2013}. A test of squeezed light at the LIGO detector was successfully performed just before the upgrade to Advanced LIGO started \cite{aasi_enhanced_2013}, and an installation of squeezed-light sources at the Advanced LIGO and Advanced Virgo detectors is currently in progress \cite{oelker_squeezed_2014,acernese_advanced_2015}.

Because squeezed states of lights rely on quantum correlations between the individual photons, they are more sensitive to optical loss than coherent states: when photons are lost, these correlations are also destroyed. Table-top experiments routinely achieve less than 10\% total loss from production to detection of the squeezed states, allowing squeeze factors in excess of \SI{10}{dB} and recently achieving \SI{15}{dB} \cite{vahlbruch_detection_2016}. Yet, transferring these high values to quantum-metrology applications remains a challenge. For example, the best squeezing in the large-scale interferometer GEO\,600 was measured to be \SI{4.4}{dB} \cite{schreiber_gravitational-wave_2017}. This stark contrast is due to a much higher optical and interferometric complexity, which brings the total loss to a value of $> 30\%$ \cite{the_ligo_scientific_collaboration_gravitational_2011,aasi_enhanced_2013,schreiber_gravitational-wave_2017}. Similarly, quantum-enhancement in biological measurements \cite{taylor_biological_2013} and magnetometry, using both atomic systems \cite{wolfgramm_squeezed-light_2010} or micro resonators \cite{li_quantum_2018}, has been limited by optical loss. 

Some optical loss -- such as absorption, scattering and polarisation mismatch -- stems from imperfect optical elements and its impact can hopefully be reduced with additional engineering work. Another source of optical loss is imperfect matching between the wavefronts of the involved light fields, i.e.\ mode-mismatch or axial misalignment. This loss channel is often difficult to control; e.g.\ in gravitational-wave detectors, sophisticated automatic alignment systems are in use \cite{morrison_automatic_1994,schreiber_alignment_2016}, as well as adaptive mode-matching with moveable lenses \cite{wittel_active_2015} and/or thermal deformation of optical elements \cite{luck_thermal_2004}. The latter two methods are, however, limited to quasi-static corrections and cannot adapt well to dynamically changing light fields.

An alternative way to alleviate the loss of squeezing due to mode-mismatch is to additionally squeeze a small number of higher-order transverse modes of light. T\"oyr\"a et al.\ \cite{toyra_multi-spatial-mode_2017} theoretically analyzed the improvement that can be achieved for squeezing the Hermite-Gaussian \TEM20 and \TEM02 modes in addition to the fundamental mode \TEM00. In their simulations, they are able to recover almost the full nonclassical sensitivity gain, for mode-mismatching loss as high as \SI{15}{\percent}.

Here we experimentally demonstrate the reduction of mode-matching loss in an interference experiment by additional squeezing of a higher-order transverse mode. Our proof-of-principle experiment almost fully compensates a (7\%) mode-mismatching loss between a squeezed field and the local-oscillator field of a balanced homodyne detector (BHD) \cite{zhang_effects_2017}. Since the underlying theory of mode-coupling is exactly the same, our results are also applicable to mode-mismatch in cavity setups.


\begin{figure}[tb]
    \includegraphics[width=\linewidth]{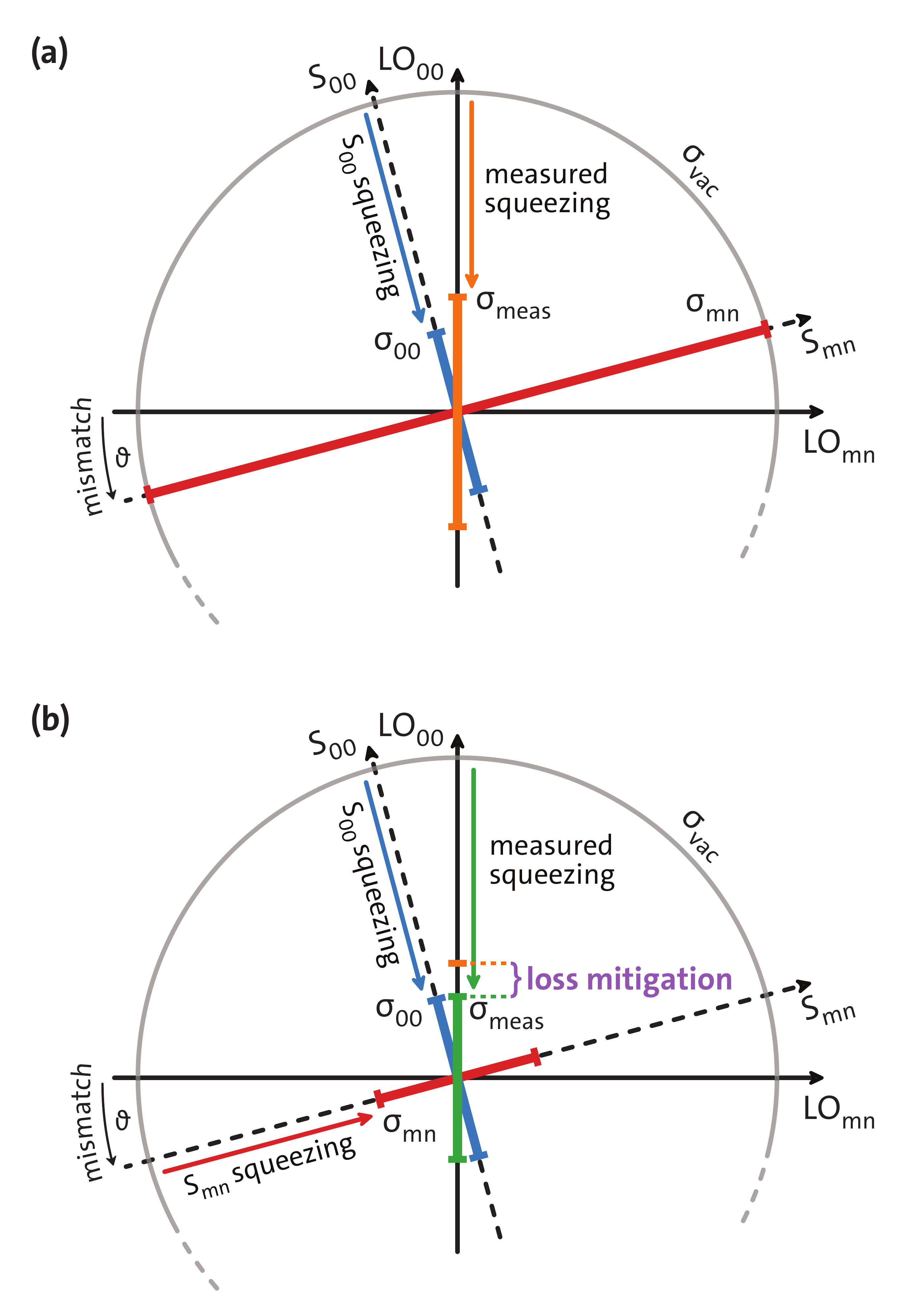}
    \caption{
    Illustration of the concept behind mode-matching loss compensation with squeezed higher-order modes. Two orthogonal axes represent the electric fields in two orthogonal transverse spatial modes. The vacuum uncertainty of any combination of modes is represented by the outer circle $\unc{\rm vac}$. Mode \Soo{} corresponds to the mode carrying the signal of interest and should ideally be aligned with the mode \Loo{} of the local oscillator in balanced-homodyne detection, which defines the mode that is actually measured. Mode-mismatch, represented by a rotation by angle $\vartheta$, then couples quantum noise $\unc{mn}$ from higher-order mode(s) into the measured quantum noise $\unc{\rm meas}$ in \Loo{}, as given by Eq.~\eqref{eq:mode_relation}. (a) When only mode \Soo{} is squeezed, and \Shom{} is in a vacuum state, the mode-mismatch leads to a significant contribution of vacuum uncertainty to $\unc{\rm meas}$. (b) When both \Soo{} and \Shom{} are squeezed (here with identical squeeze factors), the optical loss due to mode mismatch is mitigated.}  
    \label{fig:concept}
\end{figure}

\textbf{Conceptual description} ---
The idea behind squeezing of higher-order spatial modes for mode-mismatch compensation can be illustrated with the help of \figref{concept} in the following way. A BHD selectively amplifies and measures all components of a signal field that are contained in the spatial mode \Loo{}, which is defined by its so-called local oscillator (LO) beam. Usually, this mode is prepared in a well-defined \TEM00 state. Since all other \TEM{m}{n} modes are orthogonal to \TEM00, they do not interfere with the LO and do not contribute to the output of the BHD.

When the signal beam is mismatched with respect to the local oscillator field, it needs to be described by its own set of transverse modes \Smn{mn}. For a small mismatch, the relation between the two sets of modes can be conceptually represented as a rotation by a mismatch angle $\vartheta$. It connects the measured noise uncertainty $\unc{\rm meas}$ to the noise uncertainties $\unc{00}$ and $\unc{mn}$ in the signal field \Soo{} and its higher-order modes \Shom, respectively, by the relation
\begin{equation}
\begin{aligned}
    \unc{\rm meas} &= \sqrt{\var{00}\cos^2\vartheta + \var{mn}\sin^2\vartheta}\,,
\end{aligned}
\label{eq:mode_relation}
\end{equation}
where the noise in the individual modes is assumed to be uncorrelated.

When \Shom{} is in a vacuum state, $\unc{mn}=\unc{\rm vac}$, this relation implies that the mismatch is equivalent to an optical loss $\epsilon = \sin^2\vartheta$ acting on the (squeezed) mode \Soo. \figref{concept}(a) visualizes this with the rotated, dashed coordinate system. The combined projection of the noise in the \Soo{} and \Shom{} modes onto the \Loo{} mode results in a measurement noise $\unc{\rm meas}$ that exceeds the original squeezed noise $\unc{00}$ in the signal beam by far. For more complex mismatches, the same considerations apply and extend to multiple mode dimensions.

From Eq.~\eqref{eq:mode_relation}, the solution becomes obvious: if the noise in \Shom{} is also squeezed, then its contribution to the measurement noise can be significantly reduced. This is visualized in \figref{concept}(b), where the squeezed $\unc{mn}$ leads to a reduction of the noise components along the \Loo{} axis. With the exemplary squeeze factors chosen here for illustration, the initial squeezing is fully restored.

Generating squeezed states of light in higher-order spatial modes was investigated in the context of measuring lateral displacement and tilt of a laser beam \cite{treps_quantum_2003}. More recently, applications of spatial squeezing in quantum imaging have been studied, see e.g. \cite{brida_experimental_2010}. Higher-order modes can also serve as additional (quantum) communication channels in quantum information networks, either increasing the channel capacity by the number of modes that are used, or representing individual modes of a multi-mode entangled state \cite{armstrong_programmable_2012}.

There are two main approaches to the production of squeezing in higher-order spatial modes: via reshaping or via direct squeezing \cite{semmler_single-mode_2016}. In the first approach, squeezing is conventionally generated in the fundamental mode, and then converted into higher-order modes with the use of phase plates \cite{treps_quantum_2003} or spatial-light modulators (SLM) \cite{semmler_single-mode_2016}. Especially the latter allows for a very flexible generation of almost arbitrary mode shapes, however the overall efficiency and optical loss is limited by the resolution of the SLM. In the direct approach, e.g.\ \cite{lassen_generation_2006}, the nonlinear parametric cavity is aligned such that it resonates on the desired higher-order mode. This approach yields very pure states that can in principle have the same amount of squeezing as the fundamental mode, however it is only viable for low mode orders and does not support arbitrary mode shapes.


\begin{figure}
    \includegraphics[width=\linewidth]{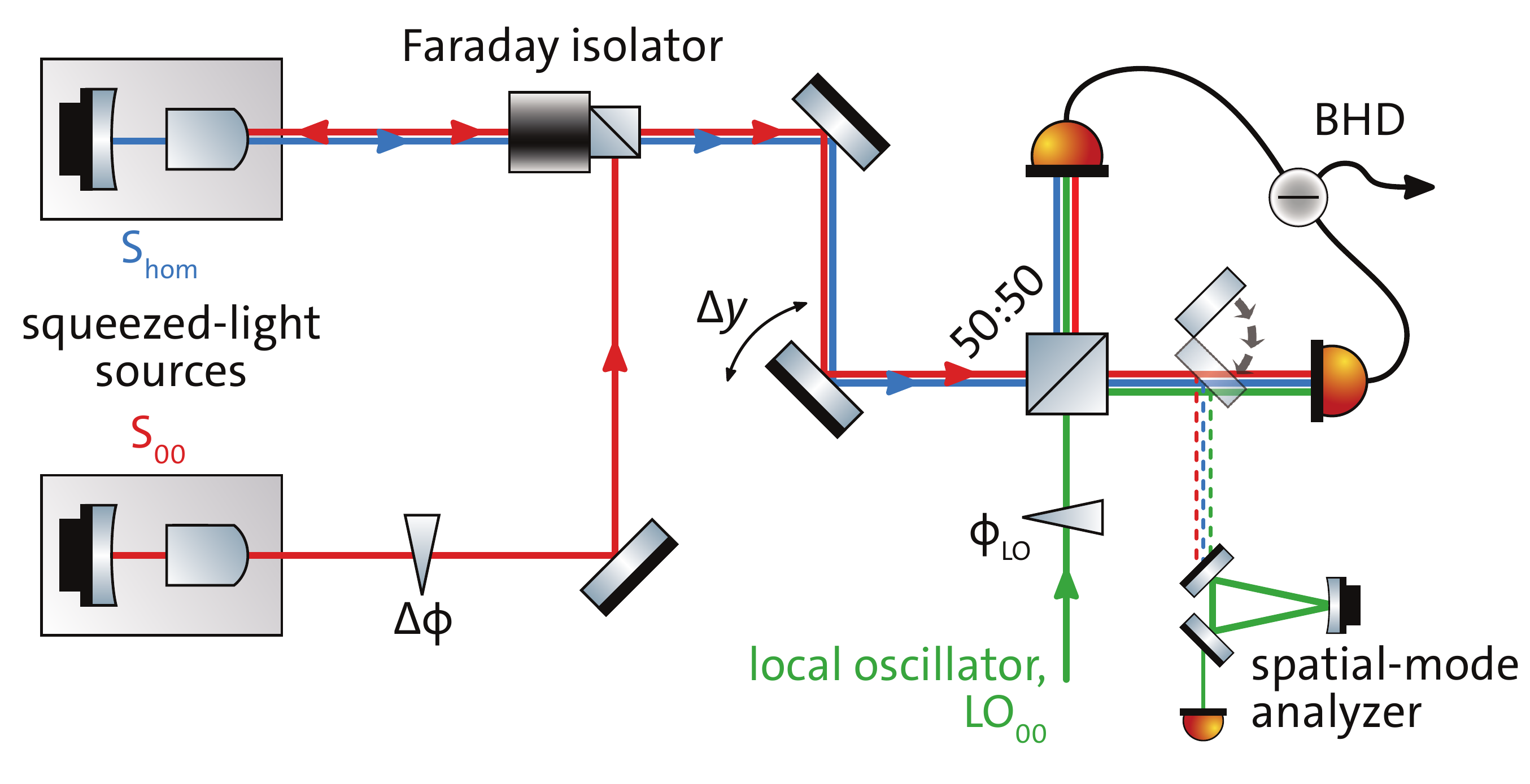}
    \caption{Schematic of our proof-of-concept of using squeezed states of light in higher-order transverse modes to compensate for optical loss caused by mode-mismatch. Squeezed-light source \Soo{} produced output states in the fundamental \TEM00 mode, while squeezed-light source \Shom{} produces output states in the \TEM01 mode. The two fields were combined with a Faraday isolator and then sent towards a balanced homodyne detector (BHD). The BHD's local oscillator \Loo{} was contained in the \TEM00 fundamental mode. Mode-matching loss was intentionally introduced by a vertical misalignment of one of the steering mirrors by $\Delta y$. The resulting mode content could be examined with a spatial-mode analyzing cavity in one of the paths of the BHD, when the two squeezing resonators were operated with a strong carrier field but without pumping.}
    \label{fig:setup}
\end{figure}

\textbf{Experimental setup} ---
Our experiment followed the direct approach, with the setup depicted in \figref{setup}. Two squeezed-light sources produced continuous-wave squeezed states of light at a wavelength of \SI{1064}{nm}, using the same optical assembly as described in detail in \cite{steinlechner_quantum_2013}. Here we label the two sources by the mode that they produced: while \Soo{} was held resonant for a \TEM00 mode, the cavity of \Shom{} was held on resonance for a higher-order mode, specifically we chose the \TEM01 mode. This was achieved by intentionally introducing a vertical displacement in the control field at \SI{1064}{nm}, which enters the cavity and is used both for locking and as an alignment reference. To obtain sufficient nonlinear gain inside the squeezing cavity, also the pump field at \SI{532}{nm} had to be vertically adjusted. For both control field and pump field, the coupling to the cavity's \TEM01 mode was about \SI{30}{\percent}, which is about three quarters of the maximally achievable mode-overlap between a \TEM01 mode and a displaced \TEM00 mode. In theory, this coupling could be as good as 43\%.

\begin{figure}
    \includegraphics[width=\linewidth]{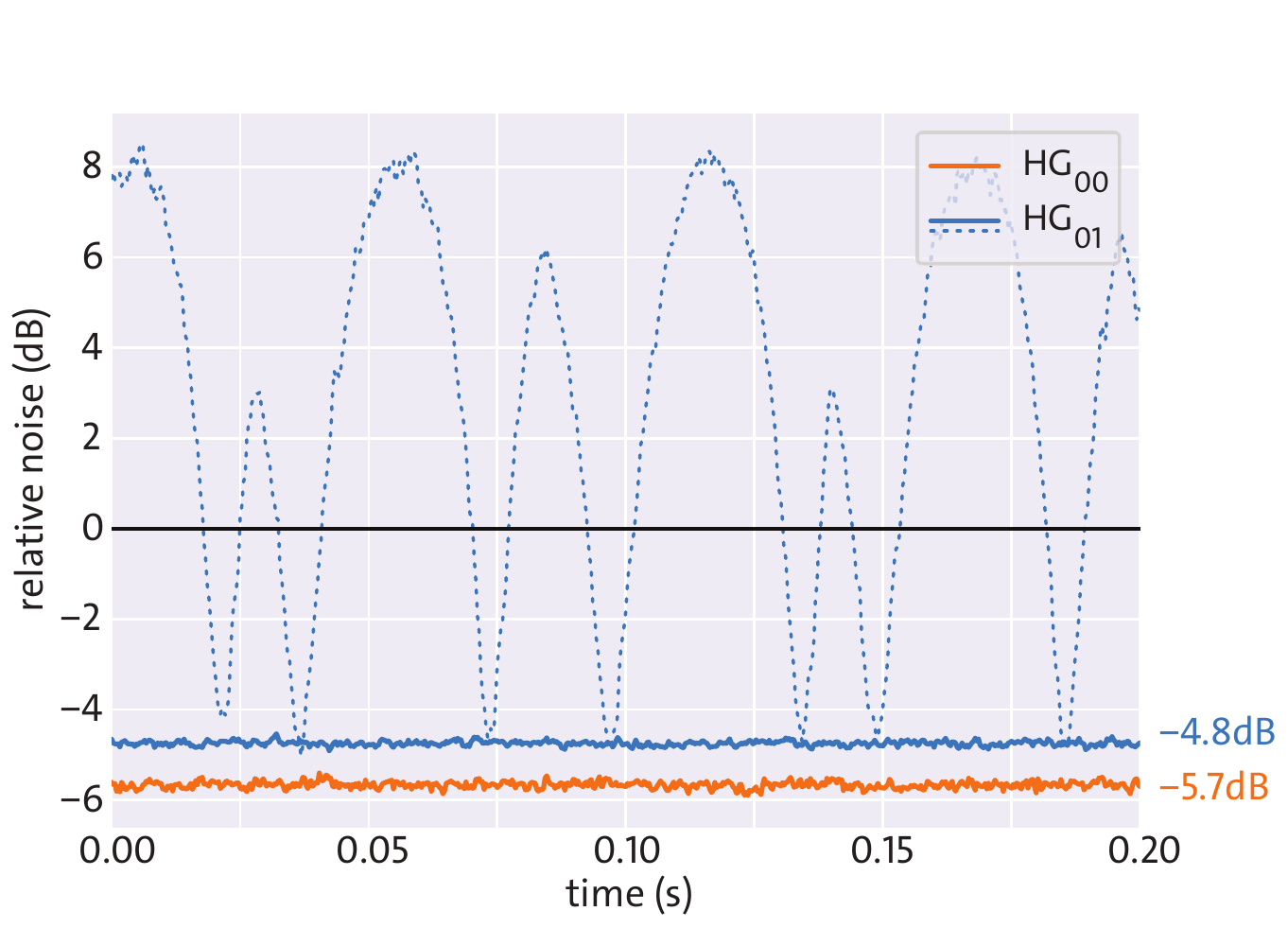}
    \caption{Characterisation of squeezed states of light from squeezing resonator \Shom{}, when resonating in the \TEM01 transverse mode, compared to resonating in the \TEM00 mode. In each case, the local oscillator was provided in the corresponding mode to achieve an optimal mode overlap at the balanced homodyne detector. The dashed blue curve was produced by scanning the local oscillator phase $\phi_{LO}$, thus showing the oscillation between squeezing and anti-squeezing. All traces were acquired with a resolution bandwidth of \SI{300}{kHz} and a video bandwidth of \SI{300}{Hz}, at a zero-span frequency of \SI{5}{MHz} and with the local oscillator power set to \SI{2.25}{mW}. Except for the scanning trace, an averaging filter of 5 was active.}
    \label{fig:tem01sqz}
\end{figure}

The squeezed field \Soo{} was first sent through a Faraday isolator, and then reflected off the \Shom{} cavity (as \Shom{} was not resonant for the \Soo{} mode) \footnote{This arrangement was chosen because of constraints of a pre-existing setup. To keep propagation loss for the \Soo{} mode as low as possible, the roles of the two squeezed-light sources should be reversed}. A piezo-mounted mirror between the two squeezers served to adjust the relative phase of the two squeezed fields $\Delta\phi$. The combined fields  \Soo{} and \Shom{} then travelled through the Faraday isolator towards a balanced homodyne detector, where they were overlapped at a 50:50 beam splitter with a strong local oscillator field. Both beam splitter outputs were detected with photodetectors and the difference in photo currents was taken. After amplification, the output power spectrum was measured with a spectrum analyzer at a sideband frequency of \SI{5}{MHz}. One of the beam-splitter outputs could be optionally sent towards a spatial-mode analyzer, which was a ring cavity specifically designed such that there was almost no mode degeneracy up to very high mode orders. This spatial-mode analyzer was used to investigate the mode content of all fields arriving at the BHD and additionally served as an alignment reference. The local oscillator field came from a mode-filtering cavity as well, and could therefore be prepared in a very pure mode state. \figref{tem01sqz} shows that our setup generated about \SI{4.8}{dB} of squeezing in the \TEM01 mode, which to our knowledge is the highest amount of squeezing in a higher-order spatial mode reported so far.

\begin{figure}
    \includegraphics[width=\linewidth]{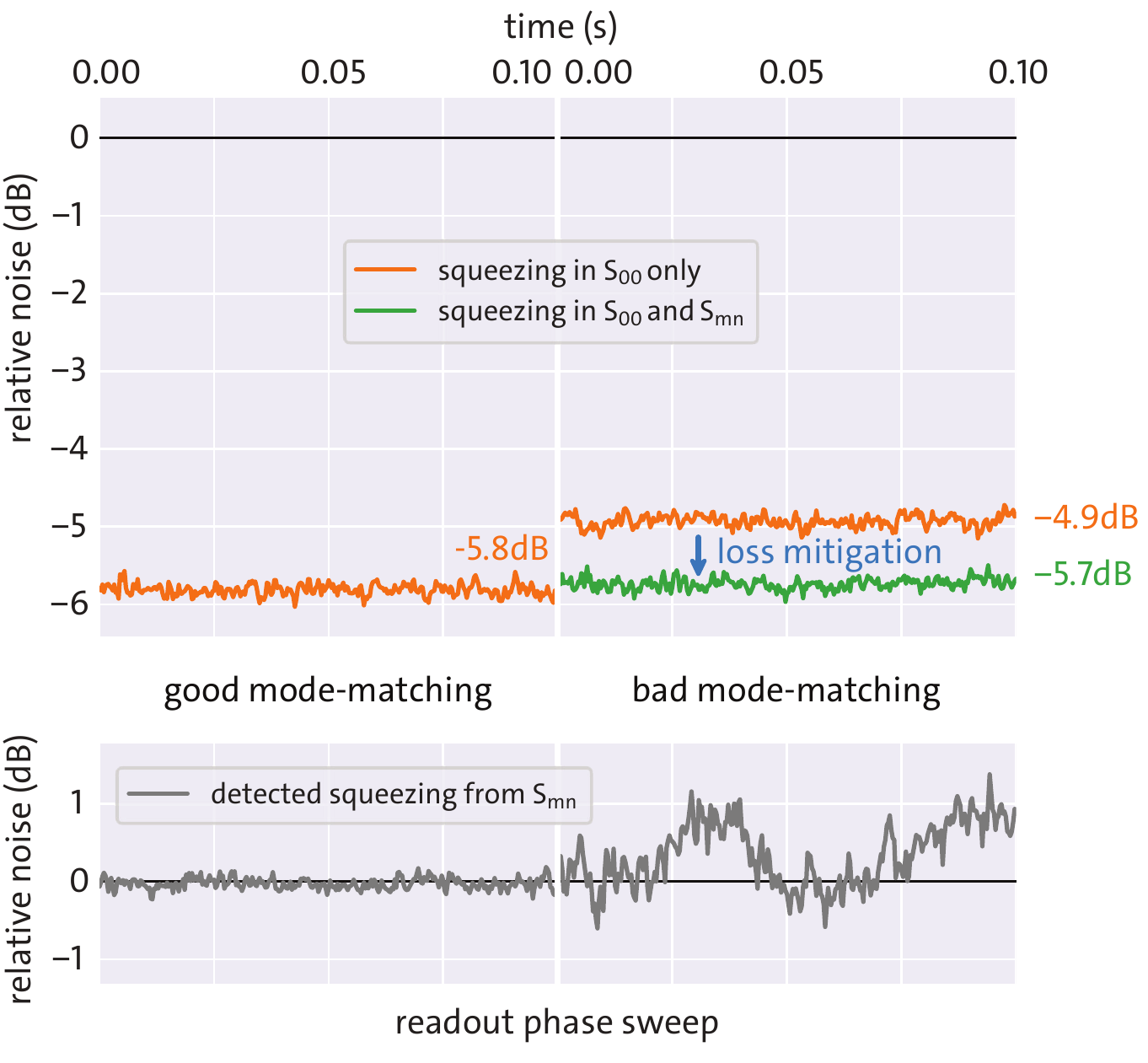}
    \caption{Confirmation of mitigating mode-matching loss via squeezed higher-order modes. Top left, \SI{5.8}{dB} of squeezing can be detected for the case where there is no vertical displacement introduced in the squeezing path, $\Delta y=0$ and thus $\Soo = \Loo$.\\
    Top right, a small vertical misalignment $\Delta y$ is introduced and the detected \Soo{} squeezing drops to \SI{4.9}{dB}. Introducing the additional squeezed-light field in \Shom{}, almost the full initial squeezing is recovered (\SI{5.7}{dB}, green trace).\\
    Bottom, detected noise during a sweep of the readout phase when only \Shom{} is squeezed. This measurement confirms that there is negligible contribution from the squeezed field in \Shom{} for the good mode-matching case, while it becomes visible in the bad mode-matching case.}
    \label{fig:new_fig_4_combined}
\end{figure}

\textbf{Results} ---
In the first step, we activated only \Shom{} and prepared the local oscillator in the \Loo{} (\TEM00) mode, with no vertical displacement $\Delta y$. In this configuration, we made sure that we were unable to detect squeezing from \Shom{}, due to the vanishing mode-overlap, see the left panel of \figref{new_fig_4_combined}. We then turned on \Soo{} as well. After reflection from \Shom{} and travelling twice through the Faraday isolator, we measured about \SI{5.8}{dB} of squeezing.

In the next step, we intentionally introduced a vertical displacement $\Delta y$ in the path between the Faraday isolator and the BHD by slightly misaligning a steering mirror in that path. The effect on the detected squeezing from \Soo{} and \Shom{}, when looked at individually, can be seen in the right part of \figref{new_fig_4_combined}. Firstly, the squeezing from \Shom{} became visible again, although only with very low squeezing values (the figure shows a sweep of the LO phase $\phi_{LO}$ of the BHD for this curve, which then also shows the anti-squeezing, making it more visible). Secondly, the detected squeezing from \Soo{} dropped from \SI{5.8}{dB} down to \SI{4.9}{dB}. From the measured squeezing to anti-squeezing for \Soo{}, we were able to estimate an additional $(7\pm 1)\%$ of loss that was introduced by the misalignment. This was also independently verified via the spatial-mode analyzer, which showed an increase in higher-order modes (mostly \TEM01 as expected) by the same 7\%. We then switched on both squeezed-light sources at the same time and carefully adjusted the relative phase between the two until we obtained the green trace in \figref{new_fig_4_combined}. This curve reached a squeezing level of \SI{5.7}{dB}, i.e.\ by combining the misaligned \Soo{} squeezing with a squeezed \Shom{} beam, we were able to recover almost all squeezing that was lost because of the misalignment. The remaining \SI{0.1}{dB} can be explained by the small amount of additional modes with $m+n>1$, which were introduced by the misalignment, but not compensated by the single additional squeezed field.

\textbf{Summary and Conclusion} ---
We have shown that mitigating optical loss from mode-mismatch between a mode carrying squeezed states and a bright local oscillator is indeed experimentally feasible by introducing additional squeezed fields in higher-order modes. Our setup used two squeezed fields in a \TEM00 and \TEM01 mode, respectively, and was able to almost completely compensate for incurred loss due to mode misalignment and brought the measured squeezing level back to within \SI{0.1}{dB} of the initial value. Although misalignment can usually be compensated quite well with auto-alignment methods, our scheme can be extended and applied to a compensation of e.g.\ wavefront-curvature mismatch by chaining together additional squeezed light sources operating in the \TEM20 and \TEM02 modes.

In measurement applications where squeezed states of light are used to increase the shot-noise limited sensitivity, a precise matching of the optical transversal mode to the instrument can be very challenging. In these situations, the setup shown here could be adapted to provide a squeezed field in a specifically tailored transversal mode configuration, to compensate for encountered loss due to mode mismatch. It is straightforward, although costly, to extend our setup to more higher-order modes, by cascading additional squeezed-light sources via Faraday isolators. Our scheme could enable much higher enhancement factors in quantum metrology, such as in squeezed-light enhanced gravitational-wave detection, where strong squeeze factors are produced but are severely degraded because of optical loss. 

\begin{acknowledgments}
This research was supported by the ERC project \emph{MassQ} (grant agreement number 339897) and the Science and Technology Facilities Council Consolidated Grant (number ST/N000633/1). D.~Töyrä was supported by the People Programme (Marie Curie Actions) of the European Union's Seventh Framework Programme FP7/2007-2013 (PEOPLE-2013-ITN) under REA grant agreement no 606176. This article has LIGO document number P1800230.
\end{acknowledgments}

\bibliography{sqhom}

\begin{thebibliography}{30}%
\makeatletter
\providecommand \@ifxundefined [1]{%
 \@ifx{#1\undefined}
}%
\providecommand \@ifnum [1]{%
 \ifnum #1\expandafter \@firstoftwo
 \else \expandafter \@secondoftwo
 \fi
}%
\providecommand \@ifx [1]{%
 \ifx #1\expandafter \@firstoftwo
 \else \expandafter \@secondoftwo
 \fi
}%
\providecommand \natexlab [1]{#1}%
\providecommand \enquote  [1]{``#1''}%
\providecommand \bibnamefont  [1]{#1}%
\providecommand \bibfnamefont [1]{#1}%
\providecommand \citenamefont [1]{#1}%
\providecommand \href@noop [0]{\@secondoftwo}%
\providecommand \href [0]{\begingroup \@sanitize@url \@href}%
\providecommand \@href[1]{\@@startlink{#1}\@@href}%
\providecommand \@@href[1]{\endgroup#1\@@endlink}%
\providecommand \@sanitize@url [0]{\catcode `\\12\catcode `\$12\catcode
  `\&12\catcode `\#12\catcode `\^12\catcode `\_12\catcode `\%12\relax}%
\providecommand \@@startlink[1]{}%
\providecommand \@@endlink[0]{}%
\providecommand \url  [0]{\begingroup\@sanitize@url \@url }%
\providecommand \@url [1]{\endgroup\@href {#1}{\urlprefix }}%
\providecommand \urlprefix  [0]{URL }%
\providecommand \Eprint [0]{\href }%
\providecommand \doibase [0]{http://dx.doi.org/}%
\providecommand \selectlanguage [0]{\@gobble}%
\providecommand \bibinfo  [0]{\@secondoftwo}%
\providecommand \bibfield  [0]{\@secondoftwo}%
\providecommand \translation [1]{[#1]}%
\providecommand \BibitemOpen [0]{}%
\providecommand \bibitemStop [0]{}%
\providecommand \bibitemNoStop [0]{.\EOS\space}%
\providecommand \EOS [0]{\spacefactor3000\relax}%
\providecommand \BibitemShut  [1]{\csname bibitem#1\endcsname}%
\let\auto@bib@innerbib\@empty
\bibitem [{\citenamefont {Schnabel}(2017)}]{schnabel_squeezed_2017}%
  \BibitemOpen
  \bibfield  {author} {\bibinfo {author} {\bibfnamefont {R.}~\bibnamefont
  {Schnabel}},\ }\href {\doibase 10.1016/j.physrep.2017.04.001} {\bibfield
  {journal} {\bibinfo  {journal} {Physics Reports}\ }\textbf {\bibinfo {volume}
  {684}},\ \bibinfo {pages} {1} (\bibinfo {year} {2017})}\BibitemShut {NoStop}%
\bibitem [{\citenamefont {Beausoleil}\ \emph {et~al.}(2003)\citenamefont
  {Beausoleil}, \citenamefont {Gustafson}, \citenamefont {Fejer}, \citenamefont
  {D’Ambrosio}, \citenamefont {Kells},\ and\ \citenamefont
  {Camp}}]{beausoleil_model_2003}%
  \BibitemOpen
  \bibfield  {author} {\bibinfo {author} {\bibfnamefont {R.~G.}\ \bibnamefont
  {Beausoleil}}, \bibinfo {author} {\bibfnamefont {E.~K.}\ \bibnamefont
  {Gustafson}}, \bibinfo {author} {\bibfnamefont {M.~M.}\ \bibnamefont
  {Fejer}}, \bibinfo {author} {\bibfnamefont {E.}~\bibnamefont {D’Ambrosio}},
  \bibinfo {author} {\bibfnamefont {W.}~\bibnamefont {Kells}}, \ and\ \bibinfo
  {author} {\bibfnamefont {J.}~\bibnamefont {Camp}},\ }\href {\doibase
  10.1364/JOSAB.20.001247} {\bibfield  {journal} {\bibinfo  {journal} {JOSA B}\
  }\textbf {\bibinfo {volume} {20}},\ \bibinfo {pages} {1247} (\bibinfo {year}
  {2003})}\BibitemShut {NoStop}%
\bibitem [{\citenamefont {Ramette}\ \emph {et~al.}(2016)\citenamefont
  {Ramette}, \citenamefont {Kasprzack}, \citenamefont {Brooks}, \citenamefont
  {Blair}, \citenamefont {Wang},\ and\ \citenamefont
  {Heintze}}]{ramette_analytical_2016}%
  \BibitemOpen
  \bibfield  {author} {\bibinfo {author} {\bibfnamefont {J.}~\bibnamefont
  {Ramette}}, \bibinfo {author} {\bibfnamefont {M.}~\bibnamefont {Kasprzack}},
  \bibinfo {author} {\bibfnamefont {A.}~\bibnamefont {Brooks}}, \bibinfo
  {author} {\bibfnamefont {C.}~\bibnamefont {Blair}}, \bibinfo {author}
  {\bibfnamefont {H.}~\bibnamefont {Wang}}, \ and\ \bibinfo {author}
  {\bibfnamefont {M.}~\bibnamefont {Heintze}},\ }\href {\doibase
  10.1364/AO.55.002619} {\bibfield  {journal} {\bibinfo  {journal} {Applied
  Optics}\ }\textbf {\bibinfo {volume} {55}},\ \bibinfo {pages} {2619}
  (\bibinfo {year} {2016})}\BibitemShut {NoStop}%
\bibitem [{\citenamefont {Vahlbruch}\ \emph {et~al.}(2008)\citenamefont
  {Vahlbruch}, \citenamefont {Mehmet}, \citenamefont {Chelkowski},
  \citenamefont {Hage}, \citenamefont {Franzen}, \citenamefont {Lastzka},
  \citenamefont {Goßler}, \citenamefont {Danzmann},\ and\ \citenamefont
  {Schnabel}}]{vahlbruch_observation_2008}%
  \BibitemOpen
  \bibfield  {author} {\bibinfo {author} {\bibfnamefont {H.}~\bibnamefont
  {Vahlbruch}}, \bibinfo {author} {\bibfnamefont {M.}~\bibnamefont {Mehmet}},
  \bibinfo {author} {\bibfnamefont {S.}~\bibnamefont {Chelkowski}}, \bibinfo
  {author} {\bibfnamefont {B.}~\bibnamefont {Hage}}, \bibinfo {author}
  {\bibfnamefont {A.}~\bibnamefont {Franzen}}, \bibinfo {author} {\bibfnamefont
  {N.}~\bibnamefont {Lastzka}}, \bibinfo {author} {\bibfnamefont
  {S.}~\bibnamefont {Goßler}}, \bibinfo {author} {\bibfnamefont
  {K.}~\bibnamefont {Danzmann}}, \ and\ \bibinfo {author} {\bibfnamefont
  {R.}~\bibnamefont {Schnabel}},\ }\href {\doibase
  10.1103/PhysRevLett.100.033602} {\bibfield  {journal} {\bibinfo  {journal}
  {Physical Review Letters}\ }\textbf {\bibinfo {volume} {100}} (\bibinfo
  {year} {2008}),\ 10.1103/PhysRevLett.100.033602}\BibitemShut {NoStop}%
\bibitem [{\citenamefont {Eberle}\ \emph {et~al.}(2010)\citenamefont {Eberle},
  \citenamefont {Steinlechner}, \citenamefont {Bauchrowitz}, \citenamefont
  {Händchen}, \citenamefont {Vahlbruch}, \citenamefont {Mehmet}, \citenamefont
  {Müller-Ebhardt},\ and\ \citenamefont {Schnabel}}]{eberle_quantum_2010}%
  \BibitemOpen
  \bibfield  {author} {\bibinfo {author} {\bibfnamefont {T.}~\bibnamefont
  {Eberle}}, \bibinfo {author} {\bibfnamefont {S.}~\bibnamefont
  {Steinlechner}}, \bibinfo {author} {\bibfnamefont {J.}~\bibnamefont
  {Bauchrowitz}}, \bibinfo {author} {\bibfnamefont {V.}~\bibnamefont
  {Händchen}}, \bibinfo {author} {\bibfnamefont {H.}~\bibnamefont
  {Vahlbruch}}, \bibinfo {author} {\bibfnamefont {M.}~\bibnamefont {Mehmet}},
  \bibinfo {author} {\bibfnamefont {H.}~\bibnamefont {Müller-Ebhardt}}, \ and\
  \bibinfo {author} {\bibfnamefont {R.}~\bibnamefont {Schnabel}},\ }\href
  {\doibase 10.1103/PhysRevLett.104.251102} {\bibfield  {journal} {\bibinfo
  {journal} {Physical Review Letters}\ }\textbf {\bibinfo {volume} {104}}
  (\bibinfo {year} {2010}),\ 10.1103/PhysRevLett.104.251102}\BibitemShut
  {NoStop}%
\bibitem [{\citenamefont {Stefszky}\ \emph {et~al.}(2012)\citenamefont
  {Stefszky}, \citenamefont {Mow-Lowry}, \citenamefont {Chua}, \citenamefont
  {Shaddock}, \citenamefont {Buchler}, \citenamefont {Vahlbruch}, \citenamefont
  {Khalaidovski}, \citenamefont {Schnabel}, \citenamefont {Lam},\ and\
  \citenamefont {McClelland}}]{stefszky_balanced_2012}%
  \BibitemOpen
  \bibfield  {author} {\bibinfo {author} {\bibfnamefont {M.~S.}\ \bibnamefont
  {Stefszky}}, \bibinfo {author} {\bibfnamefont {C.~M.}\ \bibnamefont
  {Mow-Lowry}}, \bibinfo {author} {\bibfnamefont {S.~S.~Y.}\ \bibnamefont
  {Chua}}, \bibinfo {author} {\bibfnamefont {D.~A.}\ \bibnamefont {Shaddock}},
  \bibinfo {author} {\bibfnamefont {B.~C.}\ \bibnamefont {Buchler}}, \bibinfo
  {author} {\bibfnamefont {H.}~\bibnamefont {Vahlbruch}}, \bibinfo {author}
  {\bibfnamefont {A.}~\bibnamefont {Khalaidovski}}, \bibinfo {author}
  {\bibfnamefont {R.}~\bibnamefont {Schnabel}}, \bibinfo {author}
  {\bibfnamefont {P.~K.}\ \bibnamefont {Lam}}, \ and\ \bibinfo {author}
  {\bibfnamefont {D.~E.}\ \bibnamefont {McClelland}},\ }\href {\doibase
  10.1088/0264-9381/29/14/145015} {\bibfield  {journal} {\bibinfo  {journal}
  {Classical and Quantum Gravity}\ }\textbf {\bibinfo {volume} {29}},\ \bibinfo
  {pages} {145015} (\bibinfo {year} {2012})}\BibitemShut {NoStop}%
\bibitem [{\citenamefont {Schönbeck}\ \emph {et~al.}(2018)\citenamefont
  {Schönbeck}, \citenamefont {Thies},\ and\ \citenamefont
  {Schnabel}}]{schonbeck_13_2018}%
  \BibitemOpen
  \bibfield  {author} {\bibinfo {author} {\bibfnamefont {A.}~\bibnamefont
  {Schönbeck}}, \bibinfo {author} {\bibfnamefont {F.}~\bibnamefont {Thies}}, \
  and\ \bibinfo {author} {\bibfnamefont {R.}~\bibnamefont {Schnabel}},\ }\href
  {\doibase 10.1364/OL.43.000110} {\bibfield  {journal} {\bibinfo  {journal}
  {Optics Letters}\ }\textbf {\bibinfo {volume} {43}},\ \bibinfo {pages} {110}
  (\bibinfo {year} {2018})}\BibitemShut {NoStop}%
\bibitem [{\citenamefont {Vahlbruch}\ \emph {et~al.}(2016)\citenamefont
  {Vahlbruch}, \citenamefont {Mehmet}, \citenamefont {Danzmann},\ and\
  \citenamefont {Schnabel}}]{vahlbruch_detection_2016}%
  \BibitemOpen
  \bibfield  {author} {\bibinfo {author} {\bibfnamefont {H.}~\bibnamefont
  {Vahlbruch}}, \bibinfo {author} {\bibfnamefont {M.}~\bibnamefont {Mehmet}},
  \bibinfo {author} {\bibfnamefont {K.}~\bibnamefont {Danzmann}}, \ and\
  \bibinfo {author} {\bibfnamefont {R.}~\bibnamefont {Schnabel}},\ }\href
  {\doibase 10.1103/PhysRevLett.117.110801} {\bibfield  {journal} {\bibinfo
  {journal} {Physical Review Letters}\ }\textbf {\bibinfo {volume} {117}},\
  \bibinfo {pages} {110801} (\bibinfo {year} {2016})}\BibitemShut {NoStop}%
\bibitem [{\citenamefont {{The LIGO Scientific
  Collaboration}}(2011)}]{the_ligo_scientific_collaboration_gravitational_2011}%
  \BibitemOpen
  \bibfield  {author} {\bibinfo {author} {\bibnamefont {{The LIGO Scientific
  Collaboration}}},\ }\href {http://dx.doi.org/10.1038/nphys2083} {\bibfield
  {journal} {\bibinfo  {journal} {Nature Physics}\ }\textbf {\bibinfo {volume}
  {7}},\ \bibinfo {pages} {962} (\bibinfo {year} {2011})}\BibitemShut {NoStop}%
\bibitem [{\citenamefont {Grote}\ \emph {et~al.}(2013)\citenamefont {Grote},
  \citenamefont {Danzmann}, \citenamefont {Dooley}, \citenamefont {Schnabel},
  \citenamefont {Slutsky},\ and\ \citenamefont {Vahlbruch}}]{grote_first_2013}%
  \BibitemOpen
  \bibfield  {author} {\bibinfo {author} {\bibfnamefont {H.}~\bibnamefont
  {Grote}}, \bibinfo {author} {\bibfnamefont {K.}~\bibnamefont {Danzmann}},
  \bibinfo {author} {\bibfnamefont {K.~L.}\ \bibnamefont {Dooley}}, \bibinfo
  {author} {\bibfnamefont {R.}~\bibnamefont {Schnabel}}, \bibinfo {author}
  {\bibfnamefont {J.}~\bibnamefont {Slutsky}}, \ and\ \bibinfo {author}
  {\bibfnamefont {H.}~\bibnamefont {Vahlbruch}},\ }\href
  {https://journals.aps.org/prl/abstract/10.1103/PhysRevLett.110.181101}
  {\bibfield  {journal} {\bibinfo  {journal} {Physical review letters}\
  }\textbf {\bibinfo {volume} {110}},\ \bibinfo {pages} {181101} (\bibinfo
  {year} {2013})}\BibitemShut {NoStop}%
\bibitem [{\citenamefont {{The LIGO Scientific
  Collaboration}}(2013)}]{aasi_enhanced_2013}%
  \BibitemOpen
  \bibfield  {author} {\bibinfo {author} {\bibnamefont {{The LIGO Scientific
  Collaboration}}},\ }\href {\doibase 10.1038/nphoton.2013.177} {\bibfield
  {journal} {\bibinfo  {journal} {Nature Photonics}\ }\textbf {\bibinfo
  {volume} {7}},\ \bibinfo {pages} {613} (\bibinfo {year} {2013})}\BibitemShut
  {NoStop}%
\bibitem [{\citenamefont {Oelker}\ \emph {et~al.}(2014)\citenamefont {Oelker},
  \citenamefont {Barsotti}, \citenamefont {Dwyer}, \citenamefont {Sigg},\ and\
  \citenamefont {Mavalvala}}]{oelker_squeezed_2014}%
  \BibitemOpen
  \bibfield  {author} {\bibinfo {author} {\bibfnamefont {E.}~\bibnamefont
  {Oelker}}, \bibinfo {author} {\bibfnamefont {L.}~\bibnamefont {Barsotti}},
  \bibinfo {author} {\bibfnamefont {S.}~\bibnamefont {Dwyer}}, \bibinfo
  {author} {\bibfnamefont {D.}~\bibnamefont {Sigg}}, \ and\ \bibinfo {author}
  {\bibfnamefont {N.}~\bibnamefont {Mavalvala}},\ }\href {\doibase
  10.1364/OE.22.021106} {\bibfield  {journal} {\bibinfo  {journal} {Optics
  Express}\ }\textbf {\bibinfo {volume} {22}},\ \bibinfo {pages} {21106}
  (\bibinfo {year} {2014})}\BibitemShut {NoStop}%
\bibitem [{\citenamefont {Acernese}\ \emph {et~al.}(2015)\citenamefont
  {Acernese} \emph {et~al.}}]{acernese_advanced_2015}%
  \BibitemOpen
  \bibfield  {author} {\bibinfo {author} {\bibfnamefont {F.}~\bibnamefont
  {Acernese}} \emph {et~al.},\ }\href {\doibase 10.1088/0264-9381/32/2/024001}
  {\bibfield  {journal} {\bibinfo  {journal} {Classical and Quantum Gravity}\
  }\textbf {\bibinfo {volume} {32}},\ \bibinfo {pages} {024001} (\bibinfo
  {year} {2015})}\BibitemShut {NoStop}%
\bibitem [{\citenamefont
  {Schreiber}(2017)}]{schreiber_gravitational-wave_2017}%
  \BibitemOpen
  \bibfield  {author} {\bibinfo {author} {\bibfnamefont {E.}~\bibnamefont
  {Schreiber}},\ }\emph {\bibinfo {title} {Gravitational-wave detection beyond
  the quantum shot-noise limit}},\ \href {https://dcc.ligo.org/P1800001/}
  {\bibinfo {type} {{PhD} {Thesis}}},\ \bibinfo  {school} {Leibniz Universität
  Hannover}, \bibinfo {address} {Hannover} (\bibinfo {year} {2017})\BibitemShut
  {NoStop}%
\bibitem [{\citenamefont {Taylor}\ \emph {et~al.}(2013)\citenamefont {Taylor},
  \citenamefont {Janousek}, \citenamefont {Daria}, \citenamefont {Knittel},
  \citenamefont {Hage}, \citenamefont {Bachor},\ and\ \citenamefont
  {Bowen}}]{taylor_biological_2013}%
  \BibitemOpen
  \bibfield  {author} {\bibinfo {author} {\bibfnamefont {M.~A.}\ \bibnamefont
  {Taylor}}, \bibinfo {author} {\bibfnamefont {J.}~\bibnamefont {Janousek}},
  \bibinfo {author} {\bibfnamefont {V.}~\bibnamefont {Daria}}, \bibinfo
  {author} {\bibfnamefont {J.}~\bibnamefont {Knittel}}, \bibinfo {author}
  {\bibfnamefont {B.}~\bibnamefont {Hage}}, \bibinfo {author} {\bibfnamefont
  {H.-A.}\ \bibnamefont {Bachor}}, \ and\ \bibinfo {author} {\bibfnamefont
  {W.~P.}\ \bibnamefont {Bowen}},\ }\href@noop {} {\bibfield  {journal}
  {\bibinfo  {journal} {Nature Photonics}\ }\textbf {\bibinfo {volume} {7}},\
  \bibinfo {pages} {229} (\bibinfo {year} {2013})}\BibitemShut {NoStop}%
\bibitem [{\citenamefont {Wolfgramm}\ \emph {et~al.}(2010)\citenamefont
  {Wolfgramm}, \citenamefont {Cerè}, \citenamefont {Beduini}, \citenamefont
  {Predojević}, \citenamefont {Koschorreck},\ and\ \citenamefont
  {Mitchell}}]{wolfgramm_squeezed-light_2010}%
  \BibitemOpen
  \bibfield  {author} {\bibinfo {author} {\bibfnamefont {F.}~\bibnamefont
  {Wolfgramm}}, \bibinfo {author} {\bibfnamefont {A.}~\bibnamefont {Cerè}},
  \bibinfo {author} {\bibfnamefont {F.~A.}\ \bibnamefont {Beduini}}, \bibinfo
  {author} {\bibfnamefont {A.}~\bibnamefont {Predojević}}, \bibinfo {author}
  {\bibfnamefont {M.}~\bibnamefont {Koschorreck}}, \ and\ \bibinfo {author}
  {\bibfnamefont {M.~W.}\ \bibnamefont {Mitchell}},\ }\href {\doibase
  10.1103/PhysRevLett.105.053601} {\bibfield  {journal} {\bibinfo  {journal}
  {Physical Review Letters}\ }\textbf {\bibinfo {volume} {105}} (\bibinfo
  {year} {2010}),\ 10.1103/PhysRevLett.105.053601}\BibitemShut {NoStop}%
\bibitem [{\citenamefont {Li}\ \emph {et~al.}(2018)\citenamefont {Li},
  \citenamefont {Bílek}, \citenamefont {Hoff}, \citenamefont {Madsen},
  \citenamefont {Forstner}, \citenamefont {Prakash}, \citenamefont
  {Schäfermeier}, \citenamefont {Gehring}, \citenamefont {Bowen},\ and\
  \citenamefont {Andersen}}]{li_quantum_2018}%
  \BibitemOpen
  \bibfield  {author} {\bibinfo {author} {\bibfnamefont {B.-B.}\ \bibnamefont
  {Li}}, \bibinfo {author} {\bibfnamefont {J.}~\bibnamefont {Bílek}}, \bibinfo
  {author} {\bibfnamefont {U.~B.}\ \bibnamefont {Hoff}}, \bibinfo {author}
  {\bibfnamefont {L.~S.}\ \bibnamefont {Madsen}}, \bibinfo {author}
  {\bibfnamefont {S.}~\bibnamefont {Forstner}}, \bibinfo {author}
  {\bibfnamefont {V.}~\bibnamefont {Prakash}}, \bibinfo {author} {\bibfnamefont
  {C.}~\bibnamefont {Schäfermeier}}, \bibinfo {author} {\bibfnamefont
  {T.}~\bibnamefont {Gehring}}, \bibinfo {author} {\bibfnamefont {W.~P.}\
  \bibnamefont {Bowen}}, \ and\ \bibinfo {author} {\bibfnamefont {U.~L.}\
  \bibnamefont {Andersen}},\ }\href {\doibase 10.1364/OPTICA.5.000850}
  {\bibfield  {journal} {\bibinfo  {journal} {Optica}\ }\textbf {\bibinfo
  {volume} {5}},\ \bibinfo {pages} {850} (\bibinfo {year} {2018})}\BibitemShut
  {NoStop}%
\bibitem [{\citenamefont {Morrison}\ \emph {et~al.}(1994)\citenamefont
  {Morrison}, \citenamefont {Meers}, \citenamefont {Robertson},\ and\
  \citenamefont {Ward}}]{morrison_automatic_1994}%
  \BibitemOpen
  \bibfield  {author} {\bibinfo {author} {\bibfnamefont {E.}~\bibnamefont
  {Morrison}}, \bibinfo {author} {\bibfnamefont {B.~J.}\ \bibnamefont {Meers}},
  \bibinfo {author} {\bibfnamefont {D.~I.}\ \bibnamefont {Robertson}}, \ and\
  \bibinfo {author} {\bibfnamefont {H.}~\bibnamefont {Ward}},\ }\href {\doibase
  10.1364/AO.33.005041} {\bibfield  {journal} {\bibinfo  {journal} {Applied
  Optics}\ }\textbf {\bibinfo {volume} {33}},\ \bibinfo {pages} {5041}
  (\bibinfo {year} {1994})}\BibitemShut {NoStop}%
\bibitem [{\citenamefont {Schreiber}\ \emph {et~al.}(2016)\citenamefont
  {Schreiber}, \citenamefont {Dooley}, \citenamefont {Vahlbruch}, \citenamefont
  {Affeldt}, \citenamefont {Bisht}, \citenamefont {Leong}, \citenamefont
  {Lough}, \citenamefont {Prijatelj}, \citenamefont {Slutsky}, \citenamefont
  {Was}, \citenamefont {Wittel}, \citenamefont {Danzmann},\ and\ \citenamefont
  {Grote}}]{schreiber_alignment_2016}%
  \BibitemOpen
  \bibfield  {author} {\bibinfo {author} {\bibfnamefont {E.}~\bibnamefont
  {Schreiber}}, \bibinfo {author} {\bibfnamefont {K.~L.}\ \bibnamefont
  {Dooley}}, \bibinfo {author} {\bibfnamefont {H.}~\bibnamefont {Vahlbruch}},
  \bibinfo {author} {\bibfnamefont {C.}~\bibnamefont {Affeldt}}, \bibinfo
  {author} {\bibfnamefont {A.}~\bibnamefont {Bisht}}, \bibinfo {author}
  {\bibfnamefont {J.~R.}\ \bibnamefont {Leong}}, \bibinfo {author}
  {\bibfnamefont {J.}~\bibnamefont {Lough}}, \bibinfo {author} {\bibfnamefont
  {M.}~\bibnamefont {Prijatelj}}, \bibinfo {author} {\bibfnamefont
  {J.}~\bibnamefont {Slutsky}}, \bibinfo {author} {\bibfnamefont
  {M.}~\bibnamefont {Was}}, \bibinfo {author} {\bibfnamefont {H.}~\bibnamefont
  {Wittel}}, \bibinfo {author} {\bibfnamefont {K.}~\bibnamefont {Danzmann}}, \
  and\ \bibinfo {author} {\bibfnamefont {H.}~\bibnamefont {Grote}},\ }\href
  {\doibase 10.1364/OE.24.000146} {\bibfield  {journal} {\bibinfo  {journal}
  {Optics Express}\ }\textbf {\bibinfo {volume} {24}},\ \bibinfo {pages} {146}
  (\bibinfo {year} {2016})}\BibitemShut {NoStop}%
\bibitem [{\citenamefont {Wittel}(2015)}]{wittel_active_2015}%
  \BibitemOpen
  \bibfield  {author} {\bibinfo {author} {\bibfnamefont {H.}~\bibnamefont
  {Wittel}},\ }\emph {\bibinfo {title} {Active and passive reduction of high
  order modes in the gravitational wave detector {GEO} 600}},\ \href@noop {}
  {Ph.D. thesis},\ \bibinfo  {school} {Hannover, Univ., Diss.} (\bibinfo {year}
  {2015})\BibitemShut {NoStop}%
\bibitem [{\citenamefont {Lück}\ \emph {et~al.}(2004)\citenamefont {Lück},
  \citenamefont {Freise}, \citenamefont {Goßler}, \citenamefont {Hild},
  \citenamefont {Kawabe},\ and\ \citenamefont {Danzmann}}]{luck_thermal_2004}%
  \BibitemOpen
  \bibfield  {author} {\bibinfo {author} {\bibfnamefont {H.}~\bibnamefont
  {Lück}}, \bibinfo {author} {\bibfnamefont {A.}~\bibnamefont {Freise}},
  \bibinfo {author} {\bibfnamefont {S.}~\bibnamefont {Goßler}}, \bibinfo
  {author} {\bibfnamefont {S.}~\bibnamefont {Hild}}, \bibinfo {author}
  {\bibfnamefont {K.}~\bibnamefont {Kawabe}}, \ and\ \bibinfo {author}
  {\bibfnamefont {K.}~\bibnamefont {Danzmann}},\ }\href {\doibase
  10.1088/0264-9381/21/5/090} {\bibfield  {journal} {\bibinfo  {journal}
  {Classical and Quantum Gravity}\ }\textbf {\bibinfo {volume} {21}},\ \bibinfo
  {pages} {S985} (\bibinfo {year} {2004})}\BibitemShut {NoStop}%
\bibitem [{\citenamefont {Töyrä}\ \emph {et~al.}(2017)\citenamefont
  {Töyrä}, \citenamefont {Brown}, \citenamefont {Davis}, \citenamefont
  {Song}, \citenamefont {Wormald}, \citenamefont {Harms}, \citenamefont
  {Miao},\ and\ \citenamefont {Freise}}]{toyra_multi-spatial-mode_2017}%
  \BibitemOpen
  \bibfield  {author} {\bibinfo {author} {\bibfnamefont {D.}~\bibnamefont
  {Töyrä}}, \bibinfo {author} {\bibfnamefont {D.~D.}\ \bibnamefont {Brown}},
  \bibinfo {author} {\bibfnamefont {M.}~\bibnamefont {Davis}}, \bibinfo
  {author} {\bibfnamefont {S.}~\bibnamefont {Song}}, \bibinfo {author}
  {\bibfnamefont {A.}~\bibnamefont {Wormald}}, \bibinfo {author} {\bibfnamefont
  {J.}~\bibnamefont {Harms}}, \bibinfo {author} {\bibfnamefont
  {H.}~\bibnamefont {Miao}}, \ and\ \bibinfo {author} {\bibfnamefont
  {A.}~\bibnamefont {Freise}},\ }\href {\doibase 10.1103/PhysRevD.96.022006}
  {\bibfield  {journal} {\bibinfo  {journal} {Physical Review D}\ }\textbf
  {\bibinfo {volume} {96}},\ \bibinfo {pages} {022006} (\bibinfo {year}
  {2017})}\BibitemShut {NoStop}%
\bibitem [{\citenamefont {Zhang}\ \emph {et~al.}(2017)\citenamefont {Zhang},
  \citenamefont {Danilishin}, \citenamefont {Steinlechner}, \citenamefont
  {Barr}, \citenamefont {Bell}, \citenamefont {Dupej}, \citenamefont {Gräf},
  \citenamefont {Hennig}, \citenamefont {Houston}, \citenamefont {Huttner},
  \citenamefont {Leavey}, \citenamefont {Pascucci}, \citenamefont {Sorazu},
  \citenamefont {Spencer}, \citenamefont {Wright}, \citenamefont {Strain},\
  and\ \citenamefont {Hild}}]{zhang_effects_2017}%
  \BibitemOpen
  \bibfield  {author} {\bibinfo {author} {\bibfnamefont {T.}~\bibnamefont
  {Zhang}}, \bibinfo {author} {\bibfnamefont {S.~L.}\ \bibnamefont
  {Danilishin}}, \bibinfo {author} {\bibfnamefont {S.}~\bibnamefont
  {Steinlechner}}, \bibinfo {author} {\bibfnamefont {B.~W.}\ \bibnamefont
  {Barr}}, \bibinfo {author} {\bibfnamefont {A.~S.}\ \bibnamefont {Bell}},
  \bibinfo {author} {\bibfnamefont {P.}~\bibnamefont {Dupej}}, \bibinfo
  {author} {\bibfnamefont {C.}~\bibnamefont {Gräf}}, \bibinfo {author}
  {\bibfnamefont {J.-S.}\ \bibnamefont {Hennig}}, \bibinfo {author}
  {\bibfnamefont {E.~A.}\ \bibnamefont {Houston}}, \bibinfo {author}
  {\bibfnamefont {S.~H.}\ \bibnamefont {Huttner}}, \bibinfo {author}
  {\bibfnamefont {S.~S.}\ \bibnamefont {Leavey}}, \bibinfo {author}
  {\bibfnamefont {D.}~\bibnamefont {Pascucci}}, \bibinfo {author}
  {\bibfnamefont {B.}~\bibnamefont {Sorazu}}, \bibinfo {author} {\bibfnamefont
  {A.}~\bibnamefont {Spencer}}, \bibinfo {author} {\bibfnamefont
  {J.}~\bibnamefont {Wright}}, \bibinfo {author} {\bibfnamefont {K.~A.}\
  \bibnamefont {Strain}}, \ and\ \bibinfo {author} {\bibfnamefont
  {S.}~\bibnamefont {Hild}},\ }\href {\doibase 10.1103/PhysRevD.95.062001}
  {\bibfield  {journal} {\bibinfo  {journal} {Physical Review D}\ }\textbf
  {\bibinfo {volume} {95}} (\bibinfo {year} {2017}),\
  10.1103/PhysRevD.95.062001}\BibitemShut {NoStop}%
\bibitem [{\citenamefont {Treps}\ \emph {et~al.}(2003)\citenamefont {Treps},
  \citenamefont {Grosse}, \citenamefont {Bowen}, \citenamefont {Fabre},
  \citenamefont {Bachor},\ and\ \citenamefont {Lam}}]{treps_quantum_2003}%
  \BibitemOpen
  \bibfield  {author} {\bibinfo {author} {\bibfnamefont {N.}~\bibnamefont
  {Treps}}, \bibinfo {author} {\bibfnamefont {N.}~\bibnamefont {Grosse}},
  \bibinfo {author} {\bibfnamefont {W.~P.}\ \bibnamefont {Bowen}}, \bibinfo
  {author} {\bibfnamefont {C.}~\bibnamefont {Fabre}}, \bibinfo {author}
  {\bibfnamefont {H.-A.}\ \bibnamefont {Bachor}}, \ and\ \bibinfo {author}
  {\bibfnamefont {P.~K.}\ \bibnamefont {Lam}},\ }\href {\doibase
  10.1126/science.1086489} {\bibfield  {journal} {\bibinfo  {journal}
  {Science}\ }\textbf {\bibinfo {volume} {301}},\ \bibinfo {pages} {940}
  (\bibinfo {year} {2003})}\BibitemShut {NoStop}%
\bibitem [{\citenamefont {Brida}\ \emph {et~al.}(2010)\citenamefont {Brida},
  \citenamefont {Genovese},\ and\ \citenamefont
  {Ruo~Berchera}}]{brida_experimental_2010}%
  \BibitemOpen
  \bibfield  {author} {\bibinfo {author} {\bibfnamefont {G.}~\bibnamefont
  {Brida}}, \bibinfo {author} {\bibfnamefont {M.}~\bibnamefont {Genovese}}, \
  and\ \bibinfo {author} {\bibfnamefont {I.}~\bibnamefont {Ruo~Berchera}},\
  }\href {\doibase 10.1038/nphoton.2010.29} {\bibfield  {journal} {\bibinfo
  {journal} {Nature Photonics}\ }\textbf {\bibinfo {volume} {4}},\ \bibinfo
  {pages} {227} (\bibinfo {year} {2010})}\BibitemShut {NoStop}%
\bibitem [{\citenamefont {Armstrong}\ \emph {et~al.}(2012)\citenamefont
  {Armstrong}, \citenamefont {Morizur}, \citenamefont {Janousek}, \citenamefont
  {Hage}, \citenamefont {Treps}, \citenamefont {Lam},\ and\ \citenamefont
  {Bachor}}]{armstrong_programmable_2012}%
  \BibitemOpen
  \bibfield  {author} {\bibinfo {author} {\bibfnamefont {S.}~\bibnamefont
  {Armstrong}}, \bibinfo {author} {\bibfnamefont {J.-F.}\ \bibnamefont
  {Morizur}}, \bibinfo {author} {\bibfnamefont {J.}~\bibnamefont {Janousek}},
  \bibinfo {author} {\bibfnamefont {B.}~\bibnamefont {Hage}}, \bibinfo {author}
  {\bibfnamefont {N.}~\bibnamefont {Treps}}, \bibinfo {author} {\bibfnamefont
  {P.~K.}\ \bibnamefont {Lam}}, \ and\ \bibinfo {author} {\bibfnamefont
  {H.-A.}\ \bibnamefont {Bachor}},\ }\href {\doibase 10.1038/ncomms2033}
  {\bibfield  {journal} {\bibinfo  {journal} {Nature Communications}\ }\textbf
  {\bibinfo {volume} {3}},\ \bibinfo {pages} {1026} (\bibinfo {year}
  {2012})}\BibitemShut {NoStop}%
\bibitem [{\citenamefont {Semmler}\ \emph {et~al.}(2016)\citenamefont
  {Semmler}, \citenamefont {Berg-Johansen}, \citenamefont {Chille},
  \citenamefont {Gabriel}, \citenamefont {Banzer}, \citenamefont {Aiello},
  \citenamefont {Marquardt},\ and\ \citenamefont
  {Leuchs}}]{semmler_single-mode_2016}%
  \BibitemOpen
  \bibfield  {author} {\bibinfo {author} {\bibfnamefont {M.}~\bibnamefont
  {Semmler}}, \bibinfo {author} {\bibfnamefont {S.}~\bibnamefont
  {Berg-Johansen}}, \bibinfo {author} {\bibfnamefont {V.}~\bibnamefont
  {Chille}}, \bibinfo {author} {\bibfnamefont {C.}~\bibnamefont {Gabriel}},
  \bibinfo {author} {\bibfnamefont {P.}~\bibnamefont {Banzer}}, \bibinfo
  {author} {\bibfnamefont {A.}~\bibnamefont {Aiello}}, \bibinfo {author}
  {\bibfnamefont {C.}~\bibnamefont {Marquardt}}, \ and\ \bibinfo {author}
  {\bibfnamefont {G.}~\bibnamefont {Leuchs}},\ }\href {\doibase
  10.1364/OE.24.007633} {\bibfield  {journal} {\bibinfo  {journal} {Optics
  Express}\ }\textbf {\bibinfo {volume} {24}},\ \bibinfo {pages} {7633}
  (\bibinfo {year} {2016})}\BibitemShut {NoStop}%
\bibitem [{\citenamefont {Lassen}\ \emph {et~al.}(2006)\citenamefont {Lassen},
  \citenamefont {Delaubert}, \citenamefont {Harb}, \citenamefont {Lam},
  \citenamefont {Treps},\ and\ \citenamefont
  {Bachor}}]{lassen_generation_2006}%
  \BibitemOpen
  \bibfield  {author} {\bibinfo {author} {\bibfnamefont {M.}~\bibnamefont
  {Lassen}}, \bibinfo {author} {\bibfnamefont {V.}~\bibnamefont {Delaubert}},
  \bibinfo {author} {\bibfnamefont {C.~C.}\ \bibnamefont {Harb}}, \bibinfo
  {author} {\bibfnamefont {P.~K.}\ \bibnamefont {Lam}}, \bibinfo {author}
  {\bibfnamefont {N.}~\bibnamefont {Treps}}, \ and\ \bibinfo {author}
  {\bibfnamefont {H.-A.}\ \bibnamefont {Bachor}},\ }\href {\doibase
  10.2971/jeos.2006.06003} {\bibfield  {journal} {\bibinfo  {journal} {Journal
  of the European Optical Society - Rapid publications}\ }\textbf {\bibinfo
  {volume} {1}} (\bibinfo {year} {2006}),\ 10.2971/jeos.2006.06003}\BibitemShut
  {NoStop}%
\bibitem [{\citenamefont {Steinlechner}(2013)}]{steinlechner_quantum_2013}%
  \BibitemOpen
  \bibfield  {author} {\bibinfo {author} {\bibfnamefont {S.}~\bibnamefont
  {Steinlechner}},\ }\emph {\bibinfo {title} {Quantum {Metrology} with
  {Squeezed} and {Entangled} {Light} for the {Detection} of {Gravitational}
  {Waves}}},\ \href@noop {} {\bibinfo {type} {{PhD} {Thesis}}},\ \bibinfo
  {school} {Leibniz Universität Hannover}, \bibinfo {address} {Hannover}
  (\bibinfo {year} {2013})\BibitemShut {NoStop}%
\bibitem [{Note1()}]{Note1}%
  \BibitemOpen
  \bibinfo {note} {This arrangement was chosen because of constraints of a
  pre-existing setup. To keep propagation loss for the \protect \ensuremath
  {\protect \text {S}_{00}}{} mode as low as possible, the roles of the two
  squeezed-light sources should be reversed}\BibitemShut {NoStop}%
\end{thebibliography}%

\end{document}